\documentclass{aa}
\input epsf
\begin{document}

\thesaurus{09(09.08.1; 11.09.1 M101; 11.19.2)}

\title{The Impact of Resolution on Observed \ion{H}{ii} Region
Properties from WFPC2 Observations of \object{M101}
\thanks{(Based on observations with
the NASA/ESA Hubble Space Telescope, obtained from the data Archive at the
Space Telescope Science Institute, which is operated by the Association of 
Universities for Research in Astronomy, Inc., under
NASA contract NAS 5-26555.)}}

\author{P.O.\,Pleuss\inst{1} \and C.H.\,Heller\inst{2} \and
        K.J.\,Fricke\inst{1}}
\institute{Universit\"ats-Sternwarte, Geismarlandstra\ss e 11,
           37083 G\"ottingen, Germany \and
           Department of Physics, Georgia Southern University,
           Statesboro, GA 30460-8031, USA}

\offprints{ppleuss@uni-sw.gwdg.de}
\date{Received 9 July 1999; accepted 7 July 2000}

\maketitle
\markboth{P.O.\,Pleuss,\, C.H.\,Heller,\, K.J.\,Fricke:
	  Resolution Effects on \ion{H}{ii} Regions}{}

\begin{abstract}

Two continuum subtracted $\mathrm{H}\alpha$ HST frames of \object{M101}
are used to determine the positions, angular sizes and absolute 
fluxes of 237 \ion{H}{ii} regions using a semi-automated technique.
From these we have constructed the luminosity and diameter distribution
functions.

We repeat this process on the images after artificially reducing the
linear resolution to that typically obtained with ground based imaging.
We find substantial differences in the luminosity function and
diameter distribution.  The measured internal properties, such as
central surface brightness and radial gradient are dominated by the
PSF at linear resolutions less than roughly 40\,pc FWHM.  
From the ground such resolutions are currently only obtainable for the
nearest galaxies.

Further support for the dominant role played by the seeing is provided by 
simple analytical models. We also study the clustering properties of
\ion{H}{ii} regions and their effect on the luminosity function by
construction of a Minimal Spanning Tree (MST). We find evidence for
two regimes of clustering of the \ion{H}{ii} regions and diffuse emission.
These intrinsic clustering properties in combination with the spatial
resolution typically obtainable from ground based observations might be
responsible for the break in the \ion{H}{ii} region luminosity function which
is usually found at $\log\!L_{\mathrm{H}\alpha}=38.6$\,erg/s, suggesting two
different regimes of star formation in late type spiral galaxies.

From the high resolution HST data we find a luminosity function
slope of $\alpha=-1.74\pm0.08$.  We also observe a flattening  at
luminosities $\log\!L<36.7$\,erg/s.  For the diameter distribution
we find a characteristic scale of $D_0=29.2$\,pc from an exponential fit.
However, a scale free power law with index $\beta=-2.84\pm0.16$ provides a 
better fit to the data.


\end{abstract}

\section{Introduction}
\label{sec1}

As tracers of recent massive star formation, \ion{H}{ii} regions are
indispensable.  The radiation emitted by the hot gas carries the
signature of the young hot stars to which it gave birth. 
Complementing the detailed spectroscopic work that has been done,
are studies on the statistical properties of \ion{H}{ii} regions.
These global investigations are aimed at gaining an
understanding of the connection between the large scale
dynamics and the star formation properties of galaxies (e.g. interactions
and mergers, bars and spirals, resonances and rings).  Most of this work has
focused on the luminosity and diameter distribution of the regions imaged
over full galactic disks in the line of $\mathrm{H}\alpha$.

A long standing difficulty with making such measurements has been in the
identification of the regions.  At low resolutions the blending together
of regions can be significant.  An initial estimate of this effect was
made by Kennicutt et al. (\cite{ken89}), who found that blending effects
became important at linear resolutions less than 300\,pc.  However, at
the luminosity limits now obtainable with modern CCDs, a study of 
resolution effects has not been made.

For the diameter distribution of \ion{H}{ii} regions Van den Bergh
(\cite{van81}) found  the functional form of an exponential law. Hodge
(\cite{hod87}) confirmed this form for a sample of spiral galaxies and
derived a dependence of characteristic size $D_0$ on absolute magnitude
of the galaxy. A wide range of of $D_0$ between 45\, and 560\,pc was
observed for the different galaxies.  Nevertheless, in other studies a
power law form fit the diameter distribution better than the exponential
function (Kennicutt \& Hodge \cite{ken80}, Elmegreen \& Salzer
\cite{elm99}).  Likewise, Knapen (\cite{kna98}) found a value of $D_0$
for M100 that is 1.5 to 2.2 times smaller than found in previous
determinations at lower resolutions.  The reliability of available 
diameter measurements is highly questionable.

A systematic study of the luminosity function (LF) of the
\ion{H}{ii} regions in disk galaxies was 
carried out by Kennicutt et al. (\cite{ken89}), using a sample of
30 spiral and irregular galaxies.  They found the LF
could be represented by a power law with index $\alpha=-2\pm0.5$.
A flattening at luminosities below $\log\!L_{\mathrm{H}\alpha} =
37$\,erg/s was found and  explained in terms of  a transition from
single ionizing stars to ionizing stellar clusters (McKee \& Williams
\cite{McK97} and Oey \& Clarke \cite{oey98}).  The LF slope was also
found to be systematically steeper for earlier Hubble types and  the
properties of regions in spiral arms differed to those between arms. 
Theoretical explanations for these variations were given by Oey \& Clarke
(\cite{oey98}), who showed that evolutionary effects and maximum number
of ionizing stars per cluster can account for such differences.
Some galaxies also exhibited a break in the LF at about
$\log\!L_{\mathrm{H}\alpha} = 39$\,erg/s with a steeper slope at the high
luminosity end.

Other surveys were carried out by Banfi et al. (\cite{ban93}) with a
sample of 22 spirals located mainly in the virgo cluster and by Delgado \&
Perez (\cite{del97}) with 27 spiral galaxies.  In the later survey 25\% 
of objects were identified as having double slope LFs.  More
detailed studies have also been performed on single objects. 
Scowen et al.(\cite{sco92}) have studied M101 and found an index of
$\alpha=-1.85\pm0.05$ for a power law fit to the LF. Rand (\cite{ran92})
examined M51, Walterbos \& Braun (\cite{wal92}) M31 and Wyder et al.
(\cite{wyd97}) M33. Also an example of a barred galaxy, NGC~7479, was studied
by Rozas et al. (\cite{roz99}), where evidence for a difference in the
properties of the regions located in the bar from that of the overall disk
were found.

Studies of the $\mathrm{H}\alpha$ emission in grand design spirals were
carried out by Cepa \& Beckman (\cite{cep89}, \cite{cep90}), Knapen et al. 
(\cite{kna93}) and continued in a series of four grand design spirals by Rozas
et al. (\cite{roz96a}, \cite{roz96b}, \cite{roz98}).
The LFs in all four cases showed a clear slope transition at
$\log\!L_{\mathrm{H}\alpha}=38.6$\,erg/s accompanied by a local maximum or
``glitch''.  In addition to the LF break, they also found a break at the same
luminosity in the surface brightness gradient of the \ion{H}{ii} regions.
The gradients of less luminous regions were about constant, whereas the
gradients of more luminous regions were steeper and growing with
luminosity.  Yet another break has been reported at this 
luminosity in the central surface brightness
(Beckman, priv. comm.). These features have been interpreted by 
Beckman et al. (\cite{bec98}) as representing a physical change
in the \ion{H}{ii} regions at this characteristic luminosity. 
Specifically, they contend that a transition from ionization to density
bounding occurs at $\log\!L_{\mathrm{H}\alpha}=38.6$\,erg/s.
They have also proposed the use of this feature as a standard candle.

Related studies of the diffuse ionized interstellar medium (DIG) 
have been carried out by Walterbos \& Braun 
(\cite{wal94}), Ferguson et al. (\cite{fer96}) and Greenawalt et al.
(\cite{gre97}). The diffuse emission was found to be spatially associated
with the \ion{H}{ii} regions and exhibited a variety of
different morphologies.  The fraction of emission coming from this
diffuse component was estimated at 20--50\% (Ferguson et al. \cite{fer96}). 
Leaking photons from \ion{H}{ii} regions, turbulent mixing and shock
ionization have been suggested as possible ionization mechanisms.

This paper presents statistics on the \ion{H}{ii} regions in
M101 from HST observations.  The very high resolution of the 
images will allow us to investigate in detail the role played
by resolution.  
We will find that the linear resolution required to perform reliable
measurements of central surface brightness, radial gradients, and diameters
is quite stringent and can be accomplished from the ground without the aid
of adaptive optics for only the nearest galaxies.  The effects of blending
on the luminosity function will also be found to be significant, with the
diffuse emission making an important contribution.

Also examined is the intrinsic clustering characteristics of the regions
and diffuse emission and the role this plays in shaping the behavior of
the blending.  Evidence for a possible break at a characteristic clustering
regime will be presented, which when combined with blending effects may
provide a mechanism for producing the LF discontinuity along with it's
relative stability in luminosity.
 
In Sec.~\ref{sec2} of this paper we present the data and reduction methods.
A description of the methods used to measure the properties of the \ion{H}{ii}
regions is given in Sec.~\ref{sec3}.  The results and comparisons to some
simple analytic models are then given in Sec.~\ref{sec4}, followed in
Sec.~\ref{sec5} by a discussion of their implications.

\section{Observations}
\label{sec2}

The galaxy studied here is the nearby grand design Sc galaxy
\object{M101}. 
It was chosen because it is nearly face on and lies at the relatively
close distance of about 7.4\,Mpc (Kelson et al. \cite{kel96}).  These
properties limit problems due to projection effects and provide for very
high linear spatial resolution.

\begin{figure*}
\center
{\centering\leavevmode\epsfxsize=1.6\columnwidth\epsfbox{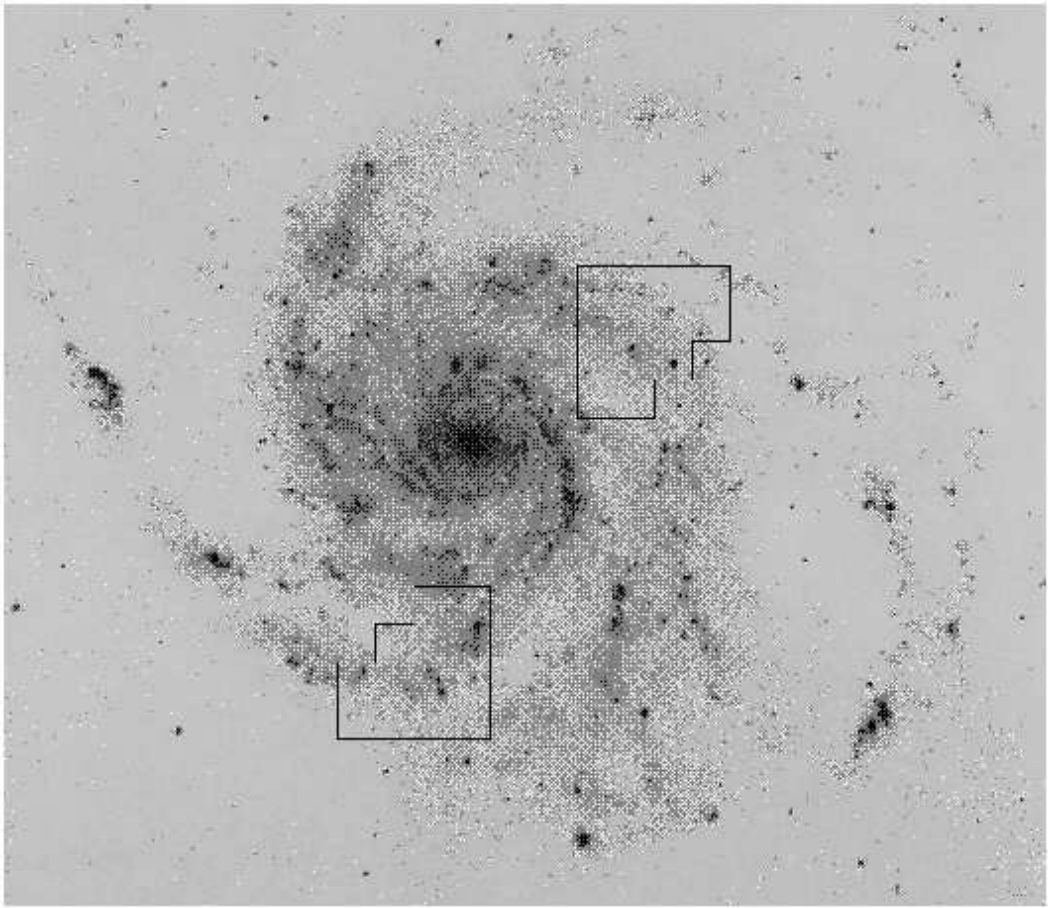}}
\caption{Ground based frame of M101 with the location of the two WFPC
frames.}
\label{fig01}
\end{figure*}

The observations of \object{M101} were obtained with the HST-WFPC2 on the
15th and 16th of September 1994 and on April 4th 1996. Both frames are
public and have been retrieved from the HST-archive. Fig.~\ref{fig01}
shows
their approximate location in the outer spiral arms. The ground based
image
was taken by Sandage (\cite{san61}) in blue light (3300--5000\,\AA) at
Mt.~Palomar.

The pixel size is $0\farcs1$ for the WF-Chips and defines the effective
resolution, i.e. the PSF FWHM is smaller. The $\mathrm{H}\alpha$
(F656N non-shifted) data consists of two sets of two exposures.  The sets
have combined exposure times of 4000\,s and 3600\,s.  Continuum
subtraction was carried out either with exposures taken with the wide
R-Filter F675W ($2\!\times\!2000$\,s) or with the medium bandwidth
V-Filter F547M ($2\!\times\!400$\,s).

Besides the HST pipeline calibration, we used the data reduction package
IRAF and especially the package STSDAS for further processing.  Particular
care was required to remove the cosmic rays through a multiple step
process. All frames were normalized to a count flux of 
$10^{-18}\,{\rm erg\,s}^{-1}\,{\rm cm}^{-2}$ using the method described in
Holtzman et al. (\cite{hol95}). Absolute fluxes were calculated using a
distance of 7.4\,Mpc.

The linear scale of the HST observations with a pixel
scale of $0\farcs1$ and distance of 7.4\,Mpc is 3.6\,pc/pixel. 
For comparison with typical ground based imaging conditions, we also
generated a set of re-pixeled frames based on a distance of 20\,Mpc and
pixel scale of $0\farcs28$. This gives a linear scale of 27\,pc/pixel 
and corresponds to a magnification factor of 7.5. 
The frames were further processed by convolving with a gaussian
corresponding to a seeing of $0\farcs8$ FWHM, leading to a resulting
linear scale of 77.6\,pc FWHM.

\begin{figure*}
\center
{\centering\leavevmode\epsfxsize=1.6\columnwidth\epsfbox{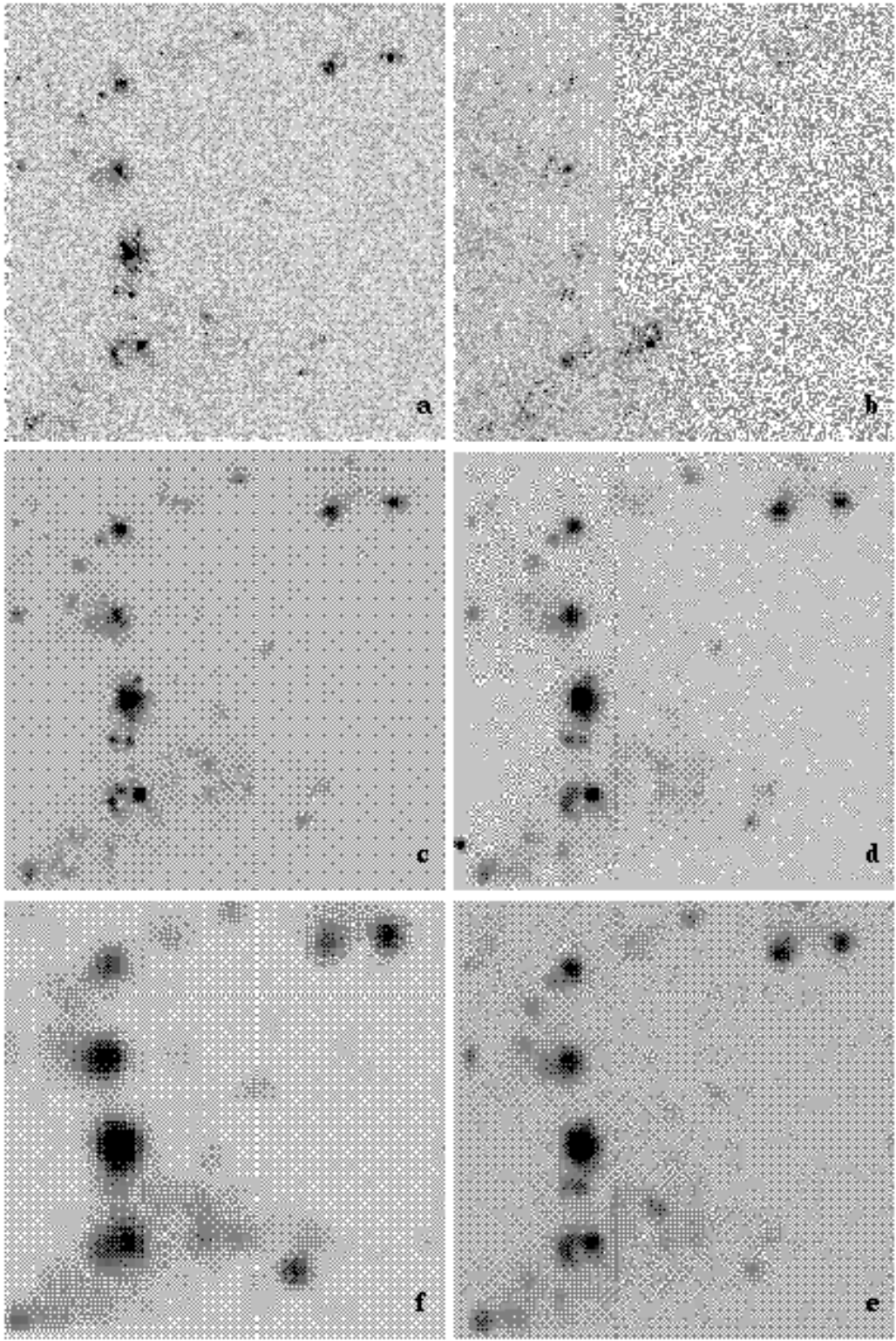}}
\caption{A visual illustration of the effects of resolution on the
	 \ion{H}{ii} region population.  The two upper panels show the
	 same area in a high resolution HST $\mathrm{H}\alpha$ ({\bf a})
         and V band ({\bf b}) image with 3.6\,pc pixelsize. The bottom
         four panels correspond, in clockwise order, to an equivalent ground
         based $\mathrm{H}\alpha$ image at distances of {\bf c} 10, {\bf d} 15,
         {\bf e} 20, and {\bf f} 35\,Mpc with an adopted seeing of
         $0\farcs8$ FWHM. The resulting linear scales are {\bf c} 38.8\,pc,
         {\bf d} 58.2\,pc, {\bf e} 77.6\,pc and {\bf f} 135.8\,pc FWHM.}
\label{fig02}
\end{figure*}

In Fig.~\ref{fig02} is an example of a high resolution $\mathrm{H}\alpha$
and V-Band image, along with several additional $\mathrm{H}\alpha$ frames
which have been shifted to varying distances using the above ground based
seeing conditions of $0\farcs8$ FWHM. The top left and right panels
correspond to the high resolution $\mathrm{H}\alpha$ and V frames,
respectively.  The lower four panels correspond, in clockwise order, to
distances of 10, 15, 20, and 35\,Mpc and linear scales of 38.8\,pc,
58.2\,pc, 77.6\,pc and 135.8\,pc FWHM, respectively.

\section{Methods}
\label{sec3}

\subsection{\ion{H}{ii}-Region Definition, High Resolution}
\label{sec3.1}

As the first selection criterion for \ion{H}{ii} regions we adopt
the requirement that they consist of at least nine contiguous pixels,
each with an intensity of at least three times the r.m.s. noise level
over the {\it local} background, which includes any contribution
from local diffuse emission. With an overall r.m.s. of about 2 normalized
counts, the minimum luminosity according to this definition is
$10^{35.6}$\,erg/s and the minimum diameter is 10.8\,pc.

At high resolution many blobs of diffuse $\mathrm{H}\alpha$ emission
with various shapes can be seen which at lower resolution are no longer
clearly discernable.  These ``false'' regions must be discriminated from
the ``true'' \ion{H}{ii} regions.
We therefore adopt a second criterion that every low surface brightness
region with clearly unsymmetric morphology must have a central stellar
counterpart in the V band. If in a multi-peaked complex we can detect more
than one well separated ionizing star cluster, it is separated into multiple
regions, otherwise the complex is treated as a single entity.

Small round high surface brightness regions without detectable ionizing
star cluster are included in the HII region definition, because we could not
exclude the possibility that a central cluster lies below our V band
detection limit or is hidden by dust. 
This definition is more precise than the one of Rand (\cite{ran92}) and
Rozas et al. (\cite{roz96a}), who define every peak of emission as an
\ion{H}{ii} region. Regions that meet the first criterion, but fail the
second one are marked as diffuse regions. Still, these account for only
$8\%$ of the detected diffuse emission.

\subsection{\ion{H}{ii} Region Definition, Low Resolution}
\label{sec3.2}

For the smeared frames we adopt the same first condition as in the 
high resolution case regarding size and flux. Therefore, with an overall
r.m.s of about 15 normalized counts the minimum luminosity is
$10^{36.4}$\,erg/s and the minimum diameter 81\,pc. This time we count
every emission peak as an \ion{H}{ii} region.

\subsection{The REGION Software}
\label{sec3.3}

To find the statistical properties of the \ion{H}{ii} regions we use the
REGION software package (Rozas et al. \cite{roz99}; Heller et al.
\cite{hel99}). This semi-automated technique identifies the \ion{H}{ii}
regions, determines their position, and measures their area and flux,
corrected for the local background. Further processing can be applied
manually: deleting, adding, separating, and the combining and rounding of
regions that do not meet our definition.

We find 237 and 162 \ion{H}{ii} regions, respectively, in the high and low
(smeared) resolution frames.  

\section{Results}
\label{sec4}

\subsection{Overall Smearing Effects}
\label{sec4.1}

A first impression of the effect of resolution can be obtained from
Fig.~\ref{fig02}, where differences between the high resolution frames and
the smeared ones in terms of the \ion{H}{ii} region sizes and morphologies
are clearly visible.

A first quantitative approach is taken by tabulating the effects that
the smearing has on the identified population:

\begin{enumerate}
\item An \ion{H}{ii} region and diffuse emission are smeared together to
      form a more luminous region (31\%).
\item More than one \ion{H}{ii} region and diffuse emission are smeared
      together to form a more luminous region (28\%).
\item Diffuse emission is reshaped by the smearing and detected as an
      \ion{H}{ii} region (41\%).
\item Small \ion{H}{ii} regions disappear in the noise.
\end{enumerate}
The percentages refer to the number of identified regions in the
smeared images.  {\it It should be noted that more than 2/3 of the 
regions in the smeared case actually consist of either multiple regions
blended together or are not genuine \ion{H}{ii} regions, in that they
lack an internal ionizing source.}

We also observed in the high resolution frames a class of compact high surface
brightness regions which show a very high degree of symmetry. They are often
found close to a large irregular region or complex. Because of their
distinct morphologies these were considered true \ion{H}{ii} regions in the
count, even if an internal ionizing source was not observed.

\subsection{Diameter Distribution}
\label{sec4.2}

\begin{figure}[t]
{\centering\leavevmode\epsfxsize=0.95\columnwidth
	   \epsfbox[100 100 450 765]{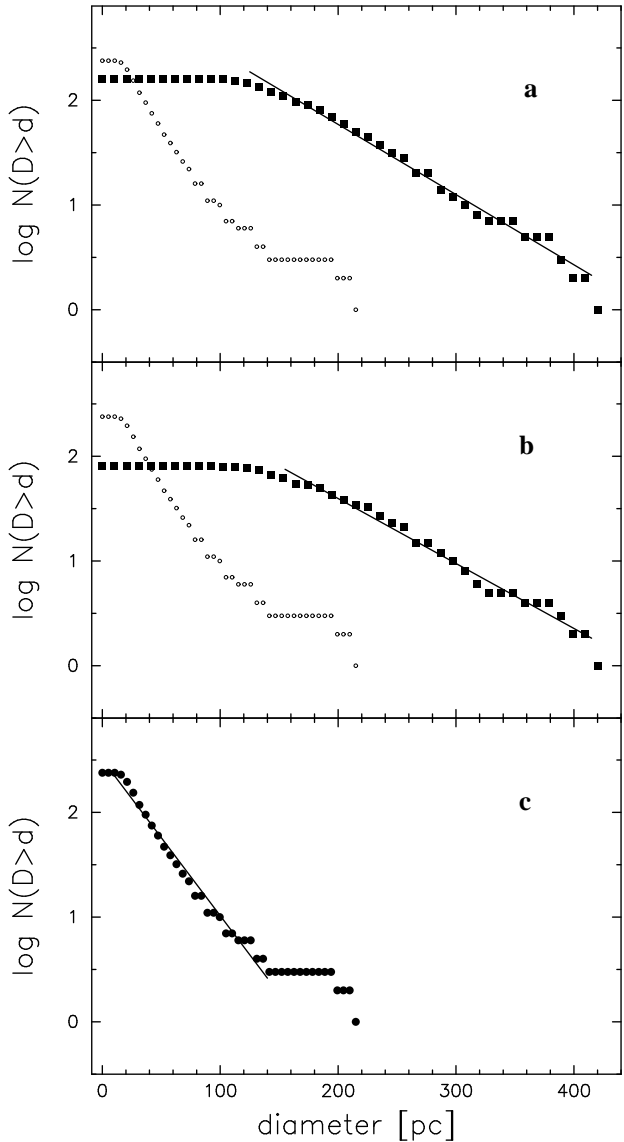}}
\caption{Integral diameter distribution. Panel {\bf a} shows the diameter
distribution for the smeared frames (filled squares, linear resolution
77.6\,pc FWHM) compared with the resolved distribution (small open circles,
linear resolution 3.6\,pc pixelsize). Panel {\bf b} shows the distribution for
the smeared frames excluding regions dominated by diffuse emission. Panel
{\bf c} gives the distribution for the resolved frames. Exponential fits are
shown with a solid line.}
\label{fig03}
\end{figure}

The integral diameter distribution shown in Fig.~\ref{fig03}
de\-mon\-strates the clear difference between the resolved and smeared
frames. The top panel shows the smeared case, the middle panel the smeared
case excluding those regions which are dominated ($>80\%$) by diffuse
emission, and the bottom panel the resolved case.
Whereas the resolved regions show a size distribution from about 10\,pc to
220\,pc, the range in the smeared case includes regions from roughly
150\,pc to greater than 400\,pc.
Parameters determined by fitting to the exponential form,
\begin{equation}
N = N_0\,{\mathrm{e}}^{-\frac{D}{D_0}},
\end{equation}
(Van den Bergh \cite{van81}) are given in Table~1.
\begin{table}
\begin{tabular}{rlrr}
{\bf } & {\bf Data} & {\bf D$_{0}$ [pc]} & {\bf N$_{0}$} \\ \hline
{\bf a} & low resolution           & 64.8 & 1285 \\
{\bf b} & low resolution corrected & 70.0 & 687 \\
{\bf c} & high resolution          & 29.2 & 318 \\
\end{tabular}
\caption{Parameters of the integral diameter distribution}
\end{table}

\begin{figure}
{\centering\leavevmode\epsfxsize=1.0\columnwidth
	   \epsfbox[60 200 500 540]{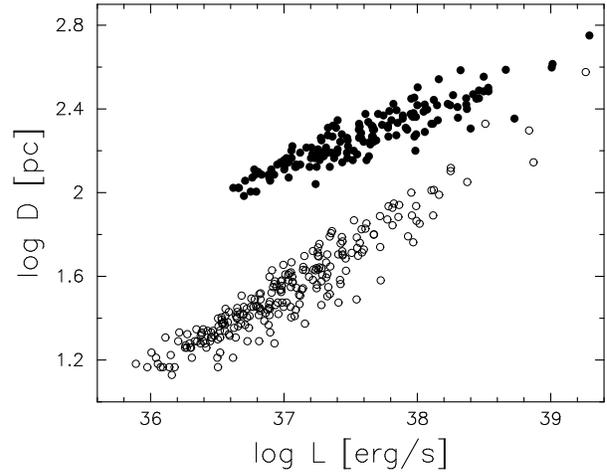}}
\caption{Logarithmic plot of region diameters against luminosity in the
         smeared (solid circles, linear resolution 77.6\,pc FWHM) and
         resolved (open circles, linear resolution 3.6\,pc pixelsize)
         cases. Both distributions can be fit by a straight line which
         represents a power law. The measured diameters of regions of all but
         the highest luminosities are significantly modified at ground based
         resolutions.}
\label{fig04}
\end{figure}

A plateau-like structure is seen in the resolved distribution at
the high luminosity end. In the high resolution study of  Wyder et al.
(\cite{wyd97}) a similar distribution is found for M33, as well as a
similar range of diameters. However, no obvious corresponding feature
is seen in the plot of $\log\!D$ vs. $\log\!L$ given in Fig.~\ref{fig04}.
The resolved and smeared data can be fit by a power law with indices
of 0.39 and 0.25, respectively. Substituting this relationship into the 
power law luminosity function of Sec.~\ref{sec4.3} gives a power law
diameter distribution with an expected index of $\beta=-2.9$.
An earlier usage of this functional form for the diameter
distribution can be found in Kennicutt \& Hodge (\cite{ken80}).
Fig.~\ref{fig04} demonstrates that the two distributions only
approach each other at rather high luminosities, indicating that 
the measured diameters of regions of all but the highest luminosities are
significantly modified at ground based resolutions.

\begin{figure}
{\centering\leavevmode\epsfxsize=1.0\columnwidth
	   \epsfbox[60 200 500 540]{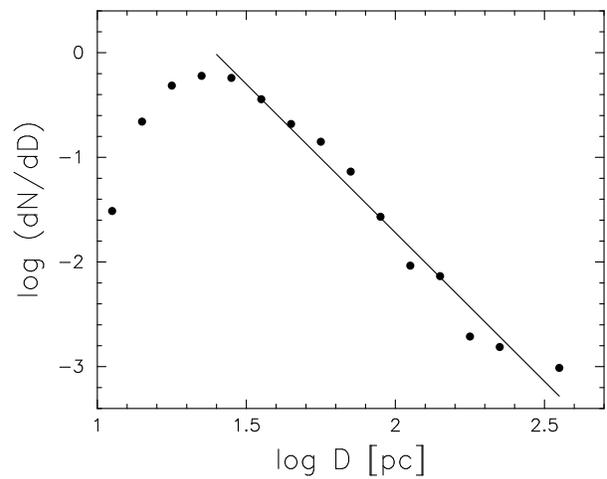}}
\caption{Differential diameter distribution along with a power law fit for the
         high resolution frames with 3.6\,pc pixelsize} 
\label{fig05}
\end{figure}

In Fig.~\ref{fig05}
we show the corresponding differential diameter distribution along with a
power law fit of the form,
\begin{equation}
{\mathrm{d}}N(D) = B\,D^\beta\,{\mathrm{d}}D,
\end{equation}
with index $\beta=-2.84\pm0.16$. The resolved data are clearly better
represented by the power law than by the exponential form. This is in
contrast to the lower resolution data, where the exponential is a better
representation.
{\it Therefore resolution effects do not only dramatically change the
characteristic diameter $D_0$, but can affect the apparent functional form
of the distribution as well.}

\subsection{Luminosity Function}
\label{sec4.3}

\begin{figure}
{\centering\leavevmode\epsfxsize=0.95\columnwidth
	   \epsfbox[100 100 450 765]{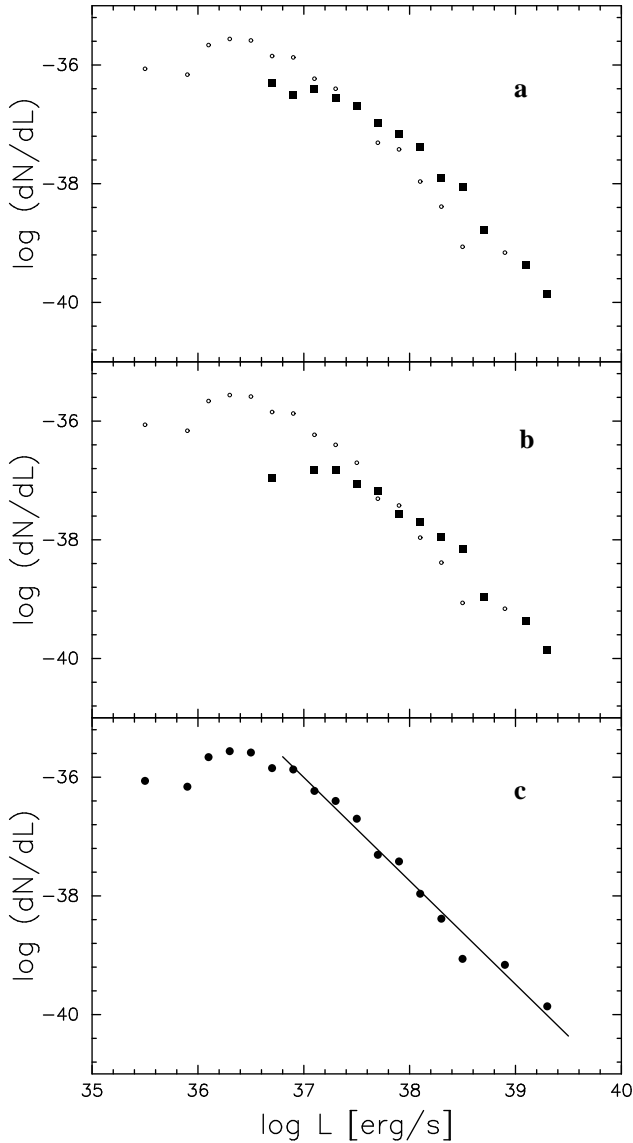}}
\caption{Differential luminosity functions. Panel {\bf a} shows the LF
of the smeared frames (full squares) in comparison with the resolved
function (open circles). Panel {\bf b} displays the LF of the
smeared frames excluding the diffuse emission dominated regions. 
Panel {\bf c} gives the LF of the resolved frames along with a power law
fit.}
\label{fig06}
\end{figure}

The combined luminosity function (LF) for the full set of resolved frames
is given in Fig.~\ref{fig06}{\bf c}.
It displays the characteristic power law form along with a flattening at
lower luminosities.  The flattening may be explained as due to a
transition from ionizing stellar clusters to single ionizing stars
(Kennicutt et al. \cite{ken89}). It should be noted that due to
completness effects most ground based studies find a turn-over instead of
a flattening.

If we include the diffuse regions in the LF, we obtain a turnover at lower
luminosities rather than a flattening although the power law index changes
little. In particular, since there are no diffuse regions with
$\log\!L>38.0$\,erg/s, the high luminosity part remains unaffected. 

Parameters of a power law fit (Kennicutt et al. \cite{ken89}),
\begin{equation}
{\mathrm{d}}N(L) = A\,L^{\alpha}\,{\mathrm{d}}L,
\end{equation}
are given in Table~2.
\begin{table}
\begin{tabular}{rlrrr}
{\bf} & {\bf LF} & {\bf Slope} & {\bf log L$_{min}$} & {\bf Regions}  \\
\hline
{\bf a} & low resolution           & -1.7  & 37.2 & 128 \\
{\bf b} & low resolution corrected & -1.6  & 37.3 &  67 \\
{\bf c} & high resolution          & -1.74 & 36.8 & 150 \\
\end{tabular}
\caption{Parameters of a power law fit to the luminosity functions}
\end{table}
The LF is consistent with a single slope with exponent
$\alpha = -1.74\,\pm{0.08}$ in the luminosity range
$\log\!L=36.7$ to 39.3\,erg/s.

The corresponding LF for the smeared frames are
given in Fig.~\ref{fig06}{\bf a}\ and\ {\bf b}.
Unlike the resolved frames, the LF for the smeared frames may be fit
consistently with either a single or double slope.  Unfortunately, 
the lack of better statistics does not allow us to make a clear 
determination of whether a break in the LF does actually occur in this
case.
However two effects can be clearly distinguished: a decrease in the number
of regions in the lower luminosity bins and an increase of regions 
in the intermediate luminosity range.  This trend becomes
even clearer when regions dominated by diffuse emission ($>$80\% in $L$) 
are excluded.  

\begin{figure}
{\centering\leavevmode\epsfxsize=1.0\columnwidth
	   \epsfbox[60 200 500 540]{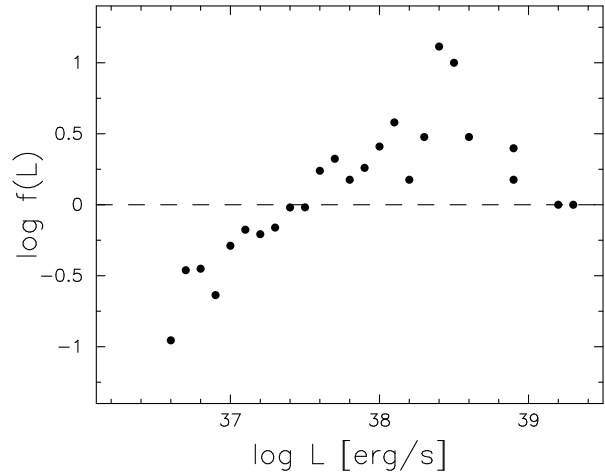}}
\caption{Logarithm of the ratio of the number of regions in
the smeared to resolved case for a given luminosity bin.}
\label{fig07}
\end{figure}

The presence of these two resolution effects is more clearly demonstrated
in Fig.~\ref{fig07}, 
where we plot the ratio of the number of regions in
the smeared to resolved case for a given luminosity bin.  In this
way one can easily see where there is an excess or deficit of regions
as compared to the resolved numbers. The figure shows a 
deficit at $\log\!L<37.6$ and an excess between 
$\log\!L=37.6$ and 38.6, which peaks at $\log\!L\approx38.4$.  
At higher luminosities the resolved and smeared LFs appear to
converge, at least within the statistical uncertainties.

These differences, combined with the additional effects
discussed in Sec.~\ref{sec4.1}, provide a picture of what happens as
the resolution is degraded. In general, regions are combined with 
diffuse emission and other regions, shifting them to
higher luminosities. While this process has been discussed
by other authors, the degree to which this occurs at typical
ground based resolutions has not been fully appreciated. 

\begin{figure}
{\centering\leavevmode\epsfxsize=1.0\columnwidth
	   \epsfbox[60 200 500 560]{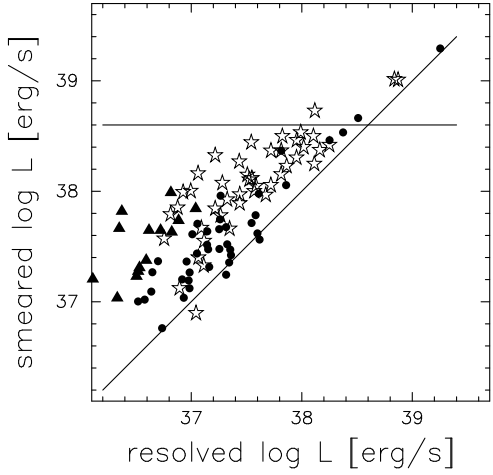}}
\caption{Region luminosity in the smeared case vs. brightest member from
resolved case. Symbols represent diffuse dominated regions (triangle), 
single regions combined with diffuse emission (dots), and blended
regions together with diffuse emission (star).}
\label{fig08}
\end{figure}

Furthermore, we propose that a break or significant reduction
in this process occurs at luminosities above $\log\!L\approx38.6$\,erg/s.
This point is demonstrated in Fig.~\ref{fig08}
where we have plotted the luminosity of regions identified in the smeared
case against their brightest member as measured from the high resolution
frames. Many regions are shifted towards a limit of $\log\!L=38.5$\,erg/s 
due to blending with other regions and diffuse emission.
This mechanism may account for a break in the LF often observed around
this limiting luminosity.

\subsection{Radial Profiles}
\label{sec4.4}

At high resolution the radial profiles typically display a core-wing
structure. The profiles in the smeared case appear gaussian, with some
core-wing structure in the brightest regions.

\subsection{Surface Brightness Gradient}
\label{sec4.5}

To compare the surface brightness gradients of the resolved regions with
earlier observations (Rozas et al. \cite{roz98}), we need to determine
a single gradient based on the profile. We decided to fit a single
gaussian to the core/wing profile and define the gradient as the slope
at the half maximum radius.

We adopt the same definition for the smeared regions, 
where it is even more appropriate. We also tried fitting a linear function
and found that the difference compared with the gaussian is
about 20\%, which is much less than the intrinsic dispersion.

\begin{figure}
{\centering\leavevmode\epsfxsize=1.0\columnwidth
	   \epsfbox[60 200 500 540]{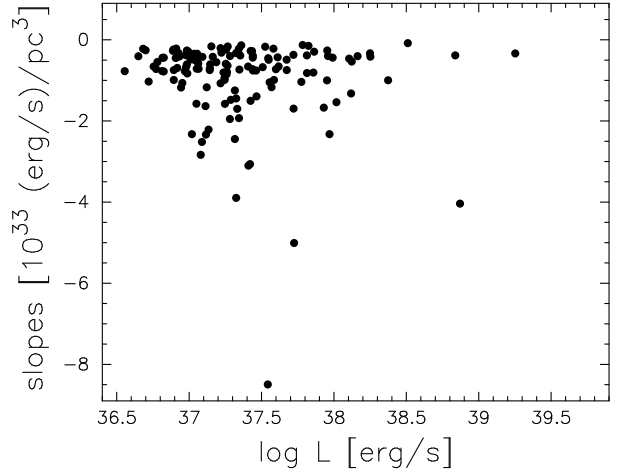}}
\caption{Surface brightness gradients of the resolved \ion{H}{ii}
regions.}
\label{fig10}
\end{figure}

The surface brightness gradients of the resolved regions are shown in
Fig.~\ref{fig10}.
In contrast to the ground based work this distribution does not show
a bilinear structure with a break at $\log\!L=38.5$\,erg/s. Rather it
exhibits a constant upper limit at all luminosities with a lot of scatter
toward steeper gradients. No other clear relation is visible.  

However, the number of regions with luminosities greater than where the
break would be is small, making the identification of a possible break
difficult. Because of this we bolster our position with a simple
analytic model of the effect of resolution on the gradients. The model
consists of 2-dimensional convolution between two gaussians, one
representing a resolved profile $F_0$, the other the point-spread function
(PSF) of the effective resolution or seeing,
\begin{eqnarray}
F_1\left(r\right) & = & F_0\left(r\right) \ast\ast
F_{\mathrm{s}}\left(r\right) \nonumber \\
& = & \frac{L}{2\pi{\sigma_{\mathrm{r}}}^2}\,
{\mathrm{e}}^{-\frac{r^2}{2{\sigma_{\mathrm{r}}}^2}} \ast\ast
\frac{1}{2\pi{\sigma_{\mathrm{s}}}^2}\,
{\mathrm{e}}^{-\frac{r^2}{2{\sigma_{\mathrm{s}}}^2}} \nonumber \\
& = &
\frac{L}{2\pi\left({\sigma_{\mathrm{r}}}^2+{\sigma_{\mathrm{s}}}^2\right)}
\,{\mathrm{e}}^{-\frac{r^2}{2\left({\sigma_{\mathrm{r}}}^2
+{\sigma_{\mathrm{s}}}^2\right)}}. 
\end{eqnarray}
The surface brightness gradients are then given by,
\begin{eqnarray}
\frac{\partial{F_0}\left(r\right)}{\partial{r}}\Bigg\vert_{r_{\mathrm{h}}} 
& = & -L\,\frac{\sqrt{2\ln{2}}}{4\pi{\sigma_{\mathrm{r}}}^3} :=
G_0\left(L,\sigma_{\mathrm{r}}\right), \label{G0} \\
\frac{\partial{F_1}\left(r\right)}{\partial{r}}\Bigg\vert_{r_{\mathrm{h}}}
& = & -L\,\frac{\sqrt{2\ln{2}}}{4\pi\left({\sigma_{\mathrm{r}}}^2
+{\sigma_{\mathrm{s}}}^2\right)^{\frac{3}{2}}} :=
G_1\left(L,\sigma_{\mathrm{r}}\right). \label{Grat}
\end{eqnarray}

A reasonable assumption for the model is to take the value for the
gradient in the resolved case, $G_0$, as a constant defined by the upper
band of points in Fig.~\ref{fig10}. Solving for $\sigma_{\mathrm{r}}^3$ in
Eq.~\ref{G0} gives,
\begin{equation}
{\sigma_{\mathrm{r}}}^3 = -\, \frac{\sqrt{2\ln{2}}}{4\pi\,G_0}\,L :=
\,k_{\mathrm{r}}\,L. 
\end{equation}
Substituting this into the smeared profile, Eq.~\ref{Grat}, we obtain
the luminosity dependent expression,
\begin{equation}
G_1\left(L\right) = -\,\frac{L\,\sqrt{2\ln{2}}}
{4\pi\left[\left({k_{\mathrm{r}}}\,L\right)^\frac{2}{3}+
{\sigma_{\mathrm{s}}}^2\right]^\frac{3}{2}}.
\end{equation}

We adopt a model resolution based on good ground 
seeing conditions of ${\rm FWHM} = 0\farcs8$ and a distance of
$D=20$\,Mpc, giving a value of $\sigma_{\mathrm{s}}=32.9$\,pc.
To find the constant $k_{\mathrm{r}}$, we fit the relation between
measured $\sigma_{\mathrm{r}}^3$ and luminosity, obtaining a value of
$k_{\mathrm{r}}= 
3.09\times{10^{-34}}\,{\mathrm{pc}}^3{\mathrm{erg}}^{-1}{\mathrm{s}}$. 

\begin{figure}
{\centering\leavevmode\epsfxsize=1.0\columnwidth
	   \epsfbox[60 200 500 540]{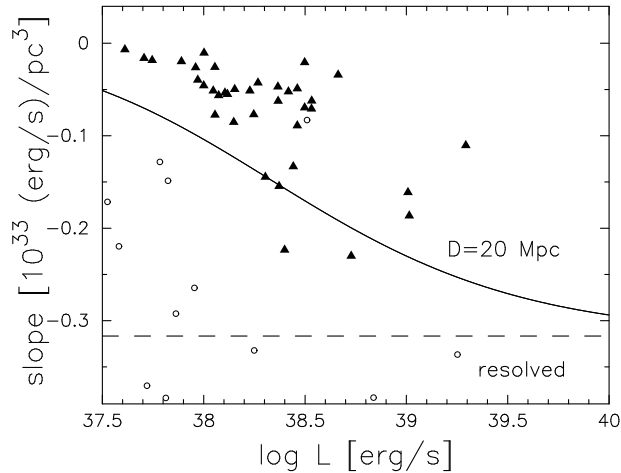}}
\caption{Surface brightness gradients. The dashed line represents the best
fit for the constant gradient of the resolved regions. The solid line is
the theoretical curve for the smeared regions. The symbols represent
measured gradients from the resolved (circles) and smeared (triangles)
frames.}
\label{fig11}
\end{figure}

The results of these calculations are plotted in Fig.~\ref{fig11},
together with the measured gradients for both resolutions.
While the statistics are lacking, the overall trend of the gradients in
the smeared case follows that found by Rozas et al. (\cite{roz98}).
They argue that this is evidence for a transition in the physical
properties of the \ion{H}{ii} regions. However, the simple model shows
that the gradients measured in the smeared frames are in fact completely
dominated by the effective seeing at the lower luminosities and do not
represent the intrinsic properties of the clouds. 
The break in the gradients which is seen in lower resolution observations
marks the zone of transition where the size of the regions is of the
same order as the effective seeing gaussian.  The fact that the
measured gradients lie above the theoretical curve is likely due to the
neglect of the contribution from diffuse emission and blending in our
simple model.

\begin{figure}
{\centering\leavevmode\epsfxsize=1.0\columnwidth
	   \epsfbox[60 200 500 540]{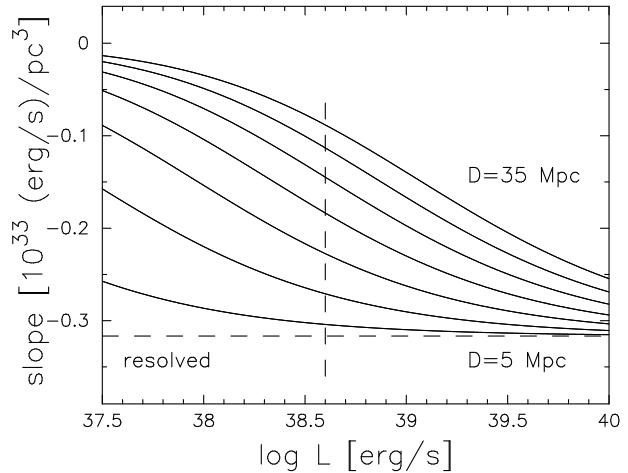}}
\caption{Theoretical curves showing the resolution effects on the 
         measured surface brightness gradients. The dashed
         line represents our constant best fit to the resolved data. Solid
         lines show the expected gradients under ground based conditions
         for distances between 5 and 35\,Mpc, corresponding to linear
         resolutions between 19.4\,pc FWHM and 135.8\,pc FWHM. The steps in
         distance are 5\,Mpc, increasing upwards.}
\label{fig12}
\end{figure}

In Fig.~\ref{fig12}
we plot various model curves representing the resolution effects on
the measured surface brightness gradients for different distances at
a seeing FWHM of $0\farcs8$.
For the given range of luminosities, all galaxies at distances larger than 
about 10\,Mpc display seeing dominated radial profiles,
corresponding to a linear resolution (FWHM) of around 40\,pc.

\subsection{Central Surface Brightness}
\label{sec4.6}

In this section we consider the central surface brightness $S_{\rm c}$ of
the \ion{H}{ii} regions. Beckman et al. (priv. comm.) find that for
regions below $\log\!L=38.6$, $S_{\rm c}$ is proportional to
$L^{1/3}$, while for regions above this value the central
surface brightness rises more steeply and shows a larger scatter.

For a theoretical treatment similar to what we have done with the
gradients, a more reliable estimate of the core profile is required. If we 
use our previous fits to the entire region we would greatly underestimate
the central brightness.  Therefore we determine $S_{\rm c}$  from a
gaussian profile that is fit only to the core area of each region in both
the resolved and smeared frames.

The core profile may be defined as
\begin{equation}
F_{\mathrm{cr}}\left(r\right) = \frac{L}{2\pi{\sigma_{\mathrm{c}}}^2}\,
{\mathrm{e}}^{-\frac{r^2}{2{\sigma_{\mathrm{c}}}^2}}
= S_{\mathrm{cr}}\,{\mathrm{e}}^{-\frac{r^2}{2{\sigma_{\mathrm{c}}}^2}}.
  \label{CSFc}
\end{equation}
Next we convolve this core profile with the seeing gaussian as
defined in Sec.~\ref{sec4.5},
\begin{eqnarray}
F_{\mathrm{cs}}\left(r\right) & = & F_{\rm cr} \ast\ast F_{\rm s}
\nonumber \\
  & = & \frac{L}{2\pi\left({\sigma_{\mathrm{c}}}^2+
       {\sigma_{\mathrm{s}}}^2\right)}\,
       {\mathrm{e}}^{-\frac{r^2}{2\left({\sigma_{\mathrm{c}}}^2+
       {\sigma_{\mathrm{s}}}^2\right)}} \nonumber \\
  & = & S_{\mathrm{cs}}\,
       {\mathrm{e}}^{-\frac{r^2}{2\left({\sigma_{\mathrm{c}}}^2+
       {\sigma_{\mathrm{s}}}^2\right)}}. \label{CSF2}
\end{eqnarray}
Substituting the approximate relationship
\begin{equation}
{\sigma_{\mathrm{c}}}^3 = \,k_{\mathrm{c}}\,L
\end{equation}
into Eqs.~\ref{CSFc} and \ref{CSF2}, we find
\begin{equation}
S_{\mathrm{cr}} = \frac{L}{2\pi{\sigma_{\mathrm{c}}}^2} = 
\frac{L^{\frac{1}{3}}}{2\pi\,k_{\mathrm{c}}^{\frac{2}{3}}}
\end{equation}
and
\begin{equation}
S_{\mathrm{cs}} = \frac{L}{2\pi\left({\sigma_{\mathrm{c}}}^2+
{\sigma_{\mathrm{s}}}^2\right)} = 
\frac{L}{2\pi\left[\left(k_{\mathrm{c}}\,L\right)^{\frac{2}{3}}+
{\sigma_{\mathrm{s}}}^2\right]}.
\end{equation}
As determined from the resolved data, we take a value for $k_{\rm c}$
of $6.3\!\times\!10^{-35}\,{\rm pc}^3{\rm erg}^{-1}{\rm s}$.

\begin{figure}
{\centering\leavevmode\epsfxsize=1.0\columnwidth
	   \epsfbox[60 200 500 540]{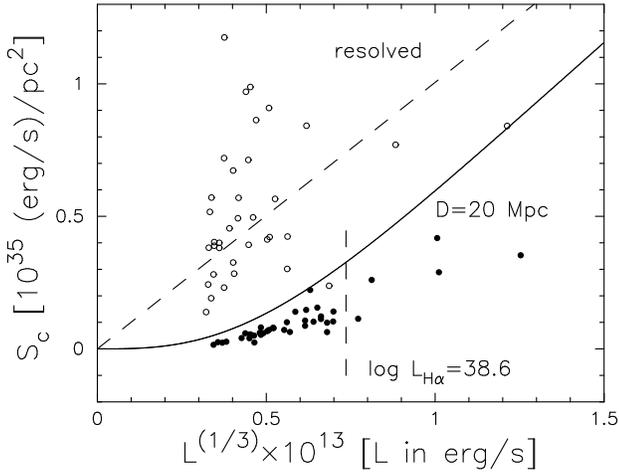}}
\caption{Central surface brightness. The dashed line represents the best
         fit of the resolved data. The solid line shows the theoretical
         curve for the smeared regions. Open circles are the resolved data
         points, solid circles represent the smeared data.}
\label{fig13}
\end{figure}

The results of the calculations are plotted against $L^{1/3}$ in
Fig.~\ref{fig13}, 
together with the data for both resolutions.

The two datasets exhibit large differences. The smeared data show a nearly
constant behavior with little scatter up to $\log\!L\approx38.6$, beyond
which $S_{\rm c}$ rises more steeply and shows larger scatter. 
This behavior closely matches that of ground based data (Beckman, priv.
comm.). In contrast, the resolved data exhibit a larger scatter at all
luminosities.

Considering the rough approximations used in deriving the theoretical
relationship for the smeared regions, the fit to the smeared data is
remarkably good.  The correspondence could be further improved if
the effects of blending and diffuse emission were taken into account.

This simple model provides a reasonable description of the
characteristics of the central surface brightness in the lower resolution
ground based observations.  {\it It demonstrates that resolution effects
are important not only for measurements of gradients, but also for 
the determination of central surface brightness.} 

\begin{figure}
{\centering\leavevmode\epsfxsize=1.0\columnwidth
	   \epsfbox[60 200 500 540]{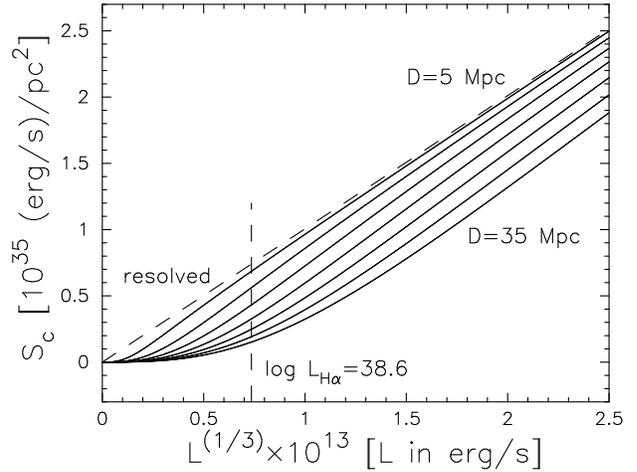}}
\caption{Theoretical curves showing the resolution effects on the 
         measured central surface brightness. The dashed line
         represents a possible fit to the resolved data. Solid lines show
         the expected central surface brightness under ground based
         conditions for distances between 5 and 35\,Mpc, corresponding
         to linear resolutions between 19.4\,pc FWHM and 135.8\,pc FWHM. The
         steps in distance are 5\,pc, decreasing upwards. The transition from a
         nearly constant value to increasing brightness is clearly visible.}
\label{fig14}
\end{figure}

In Fig.~\ref{fig14}
we plot various model curves representing the resolution effects
on the measured central surface brightness for different distances
at a seeing FWHM of $0\farcs8$.  Again we find that for galaxies at
distances greater than about 10\,Mpc the central surface brightness is
dominated by the seeing. This corresponds also to a linear scale of about
40\,pc (FWHM).

\subsection{Minimal Spanning Tree}
\label{sec4.7}

In order to investigate the clustering properties of the regions we
compute a Minimal Spanning Tree (MST) based on the spatial locations of
the \ion{H}{ii} and diffuse regions (Sec.~\ref{sec3.2}). 
The MST for any given set of points consists of a unique set of edges 
connecting the points, such that the sum of the edge-lengths
is a minimum. The MST provides a reproducible and unbiased 
graph that contains information on the intrinsic linear associations in
the point set. For the analysis of linear features the MST has been
found to be superior to other methods, such as the two point correlation
function (Barrow et al. \cite{bar85}). 

\begin{figure*}
\center
{\centering\leavevmode\epsfxsize=2.0\columnwidth
           \epsfbox[20 220 580 750]{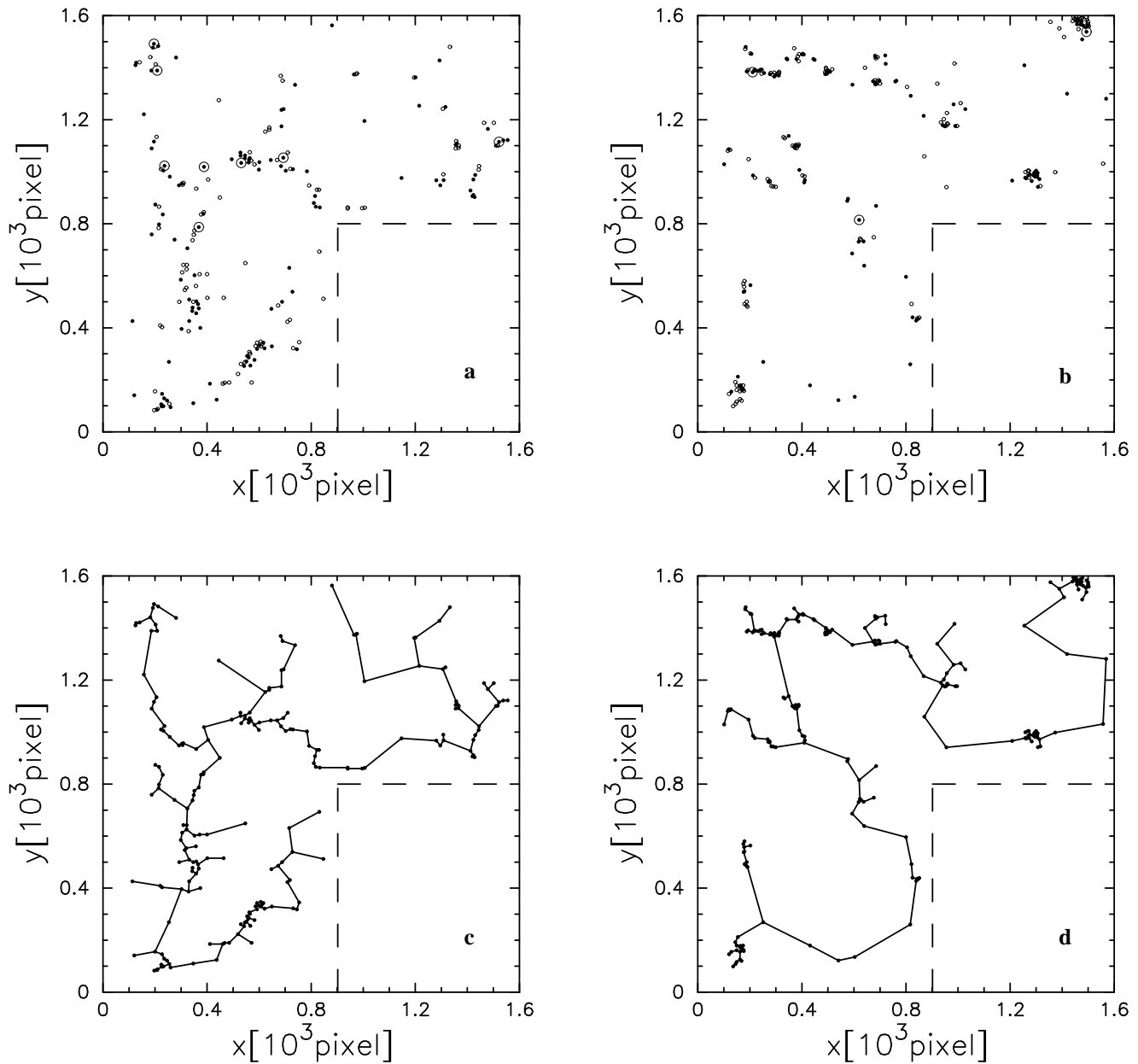}}
\caption{{\bf a} and {\bf b} Region spatial distribution for the two
         fields. The different symbols depict the \ion{H}{ii} regions
         (solid circles), diffuse regions (open circles), and high
         luminosity \ion{H}{ii} regions with
         $L_{\mathrm{H}\alpha}>10^{38}$\,erg/s (dotted circles). {\bf c} and
         {\bf d} Corresponding MSTs. Empty region at bottom right of each
         panel represents location of PC chip.}
\label{fig15}
\end{figure*}

We have chosen to include the diffuse regions in the analysis because
they are a good tracer of the spatial distribution of the diffuse emission
in general, even though they represent only 8\% of the total diffuse flux.
A graphical depiction of the region spatial distribution
and the corresponding MST are shown in Fig.~\ref{fig15} for the two
different WFPC2 fields in the galaxy.

The spatial distributions exhibit some clear features. The regions
are not located uniformly over the frames, but rather lie along ring- or
bubble-like structures. Furthermore, on the first frame four particularly
luminous regions lie along a common edge of two rings. 
An explanation for this distribution could be induced star formation in an
expanding supernovae shell.  A rough estimate of the radii of these
structures is 0.7--1.0\,kpc. This is of the order of the largest HI holes
observed in galaxies, the so-called ``superbubbles'' (Kamphuis et al.
\cite{kam91}; Oey \& Clarke \cite{oey97}).
Another possible formation mechanism  for the observed features might
be through gravitational and thermal instabilities in a differentially
rotating disk (Wada \& Norman \cite{wad99}).

Similar distribution features are also observed for HI gas. A study of 11
nearby spirals by Braun (\cite{bra97}) has found that between 60--90\% of
the total HI line flux comes from a network of filamentary features with
typical widths of about 150\,pc, which is comparable to the region
distribution in Fig.~\ref{fig15}.

\subsubsection{Resolution effects}
\label{sec4.7.1}

To model the resolution effects on \ion{H}{ii} regions, we adopt a
Gaussian profile for the regions similar to Sec.~\ref{sec4.5}. For
simplicity of the calculations we discuss the effect on two identical
region profiles.
A simple calculation shows that the peaks of two identical gaussian
profiles remain resolvable under convolution with a seeing gaussian
if the peaks are originally separated by a distance larger than
$2 \sqrt{\sigma_{\rm s}^2+\sigma_{\rm r}^2}$.
Only for the most luminous regions is $\sigma_{\rm r}$ greater than
$\sigma_{\rm s}$. Because the luminous regions are not clustered close
together (see Fig.~\ref{fig15}), we can as a resonable approximation drop
$\sigma_{\rm r}$. Therefore regions that are closer than two seeing
$\sigma_{\rm s}$ can not be separated.
The pixelsize places further constraints on the resolvability of regions.
In order to discriminate two emission peaks from each other, they must be
separated by at least one pixel of lower flux. Hence the separation
between the peaks must be at least two pixels.
Our adopted ground based conditions are given by a pixelsize and seeing
FWHM of respectively, $0\farcs28$ and $0\farcs8$, which correspond to
$\sigma_{\rm s} = 0\farcs34$. Then the limiting quantity is 
$2\sigma_{\rm s}$ of $0\farcs68$, corresponding to a linear dimension of
about 66\,pc at a distance of 20\,Mpc.

\begin{figure}
{\centering\leavevmode\epsfxsize=1.0\columnwidth
	   \epsfbox[60 200 500 540]{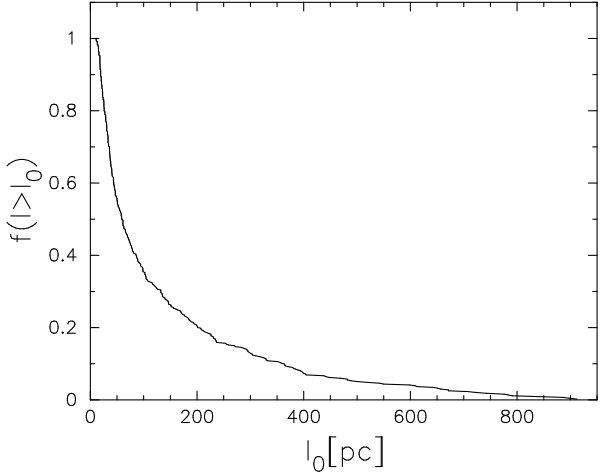}}
\caption{Integral MST edge-length distribution of the two fields combined
         at a linear pixel scale of 3.6\,pc.}
\label{fig16}
\end{figure}

The integral edge-length distribution, shown in Fig.~\ref{fig16},
gives directly the fraction of edges which have lengths above 
a given value. We can see that about 45--50\% of all
edges cannot be resolved at our adopted ground based resolution,
demonstrating the importance of blending. The degree of blending will
depend not only on the resolution, but also on the clustering properties
of the \ion{H}{ii} regions and diffuse emission.

\begin{figure}
{\centering\leavevmode\epsfxsize=0.95\columnwidth
	   \epsfbox[100 100 450 765]{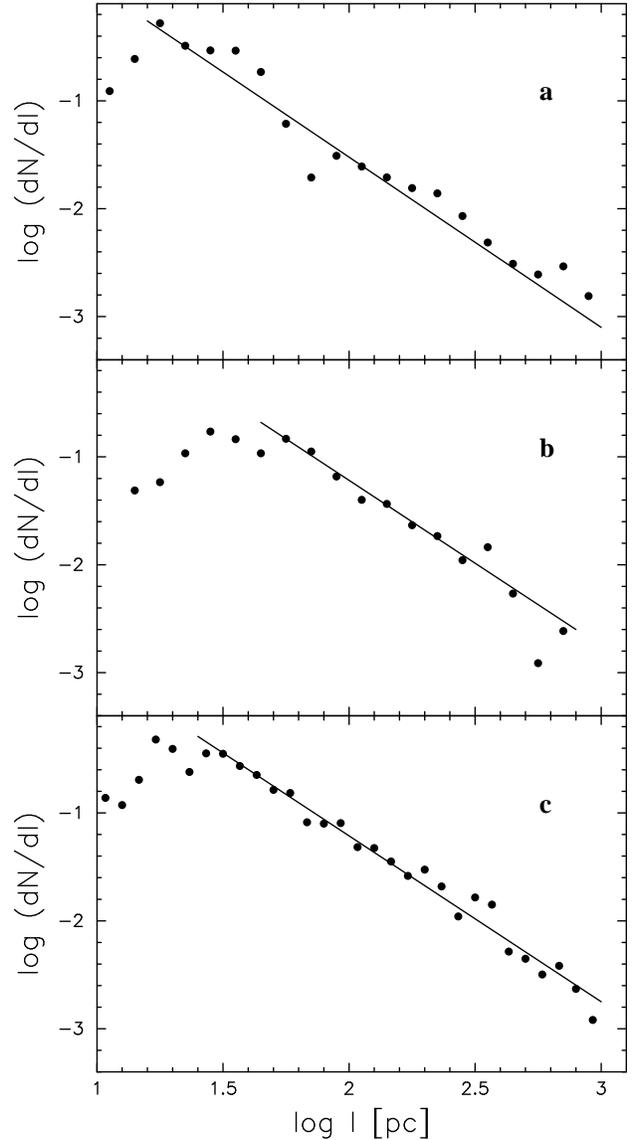}}
\caption{Differential edge-length distribution for the MST of
         \ion{H}{ii} and diffuse regions for the individual fields
         ({\bf a} \& {\bf b}) and for both fields combined ({\bf c})
         at a linear pixel scale of 3.6\,pc.}
\label{fig17}
\end{figure}

In Fig.~\ref{fig17}
we present the differential MST edge-length distribution along with
power law fits.  The parameters for the fits are given in Table~3.
\begin{table}
\begin{tabular}{lrrr}
{\bf field} & {\bf Slope} & {\bf log l$_{\rm min}$} & 
{\bf l$_{\rm min}$ [pc]} \\ \hline
{\bf a} first           & -1.49 & 1.20 & 16\\
{\bf b} second          & -1.69 & 1.65 & 45\\
{\bf c} combined        & -1.53 & 1.40 & 25\\
\end{tabular}
\caption{Parameters of power law fit to the edge-length distributions.}
\end{table}
The power law index varies little between the frames (1.5--1.7),
but the critical length where the distribution flattens 
differs substantially (16--45\,pc). This could be a hint for a
different minimum clustering regime for the two regions in the galaxy.
This range of values is below the resolution of 66\,pc of our smeared
frames.

To further quantify the impact of region clustering combined with
insufficient resolution, we simulate the blending of regions by removing
the edges that are longer than twice the seeing $\sigma_{\rm s}$. The
clipping length $l_{\rm c}$ is now treated as a function of the distance D,
\begin{equation}
l_{\rm c} = \frac{\pi\cdot0\farcs68}{3600\cdot180}\,D. \label{dtoD}
\end{equation}
We define a cluster as two or more regions which remain connected by edges 
following the clipping. With this definition all the regions in a
given cluster would be smeared together into a single region at the
specified seeing and distance.

\begin{figure}
{\centering\leavevmode\epsfxsize=1.0\columnwidth
	   \epsfbox[60 180 500 580]{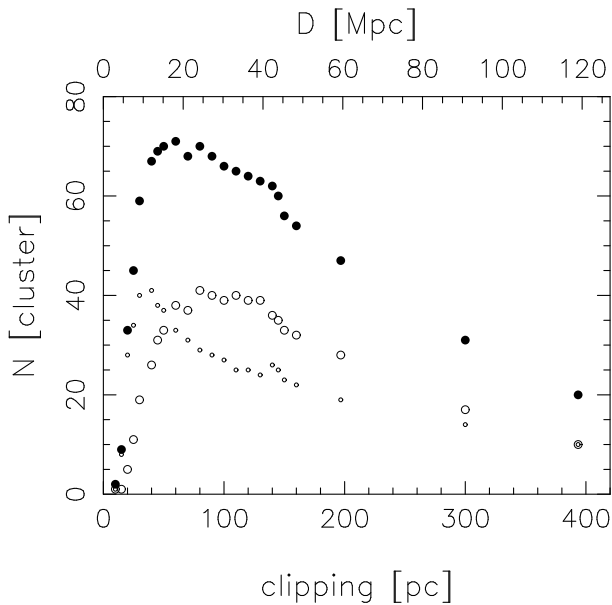}}
\caption{Number of clusters vs. clipping distance. 
         The symbols represent the clusters from the individual fields
         (open circles) and the fields combined (solid circles).  Also
         given along the top is the equivalent galaxy distance.}
\label{fig18}
\end{figure}

\begin{figure}
{\centering\leavevmode\epsfxsize=1.0\columnwidth
	   \epsfbox[60 180 500 580]{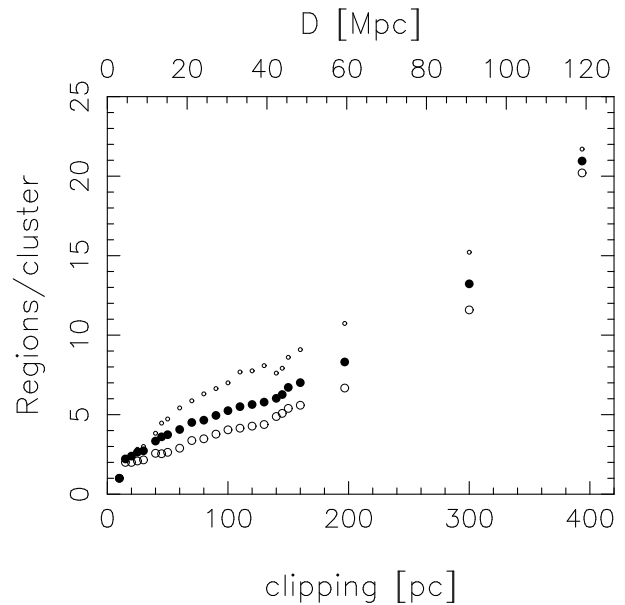}}
\caption{Average number of regions per cluster vs. clipping distance.
         The symbols represent the clusters from the individual fields
         (open circles) and the fields combined (solid circles).  Also
         given along the top is the equivalent galaxy distance.}
\label{fig19}
\end{figure}

Fig.~\ref{fig18}
shows the number of clusters as a function of the clipping distance and
the equivalent galaxy distance, with an assumed seeing of $0\farcs8$.
The curves first increase rapidly, reach a maximum number of clusters
at around a clipping length of 40\,pc, and then slowly decrease as the
clipping becomes larger.  In Fig.~\ref{fig19}
the corresponding average number of regions per cluster is given.
While differing in detail, the overall clustering characteristics of
the two fields are similar, growing approximately linearly with the
clipping (smearing). The difference in slopes could be statistical, but
also may have a physical origin in differing clustering properties of the
two fields.

The clustering behavior may be interpreted in terms of two physical
spatial regimes. Regions that are separated by edges lying in the first regime
of 40\,pc are thrown together by the smearing, creating a large number of
clusters. This first regime corresponds to the flat part of the
differential edge-length distribution. With increasing cutoff length the
clusters, which are now grouped in the second regime, are slowly blended
together according to their power law distribution. This behavior can been
seen by eye in Fig.~\ref{fig02}.

A check may be made by determining the sum of the numbers of clusters and
unclustered regions with $\log\!L>36.7$\,erg/s at a clipping of 60\,pc.
This sum should be roughly equal to the number of regions in the
smeared frames. This is because a clipping of 60\,pc corresponds to a
distance of approximately 20\,Mpc (Eqn.~\ref{dtoD}) and the smallest
detected regions in the smeared case have luminosities of
$\log\!L=36.7$\,erg/s. The totals are in fact quite close, with values of
162 and 157, respectively.

If the blending of regions were responsible for the observed break in
the luminosity function, with these clustering properties one would expect
that the break would first appear at distances of around 10\,Mpc
($0\farcs8$ FWHM) and then only change slowly with increasing distances.
The corresponding linear scale again is about $40$\,pc FWHM.

\subsubsection{Physical Clustering}
\label{sec4.7.2}

\begin{figure}
{\centering\leavevmode\epsfxsize=0.95\columnwidth
	   \epsfbox[100 300 450 765]{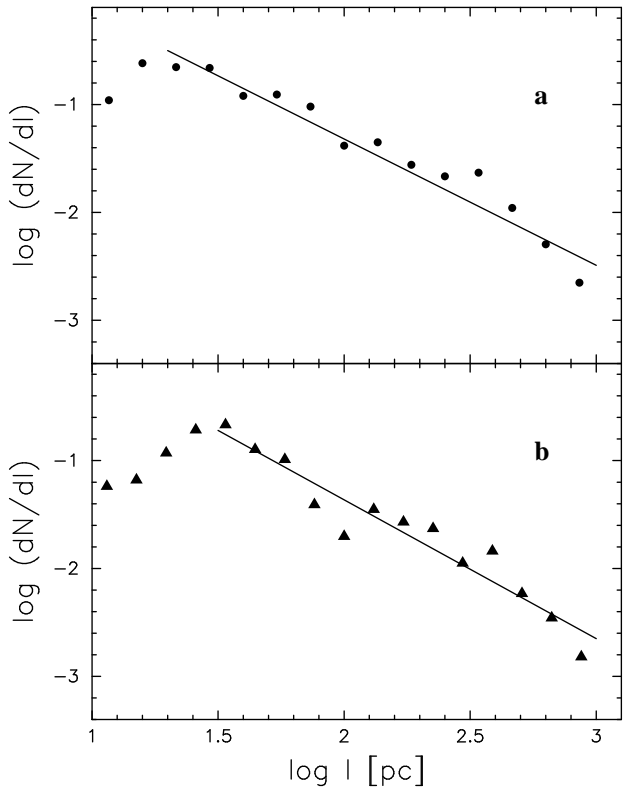}}
\caption{Differential MST edge-length distribution of {\bf a}
         \ion{H}{ii} and {\bf b} diffuse regions from both fields
         combined at a linear pixel scale of 3.6\,pc.}
\label{fig20}
\end{figure}

Putting resolution effects aside, we now examine the physical spatial
clustering properties of the \ion{H}{ii} and diffuse regions.  In
Fig.~\ref{fig20}
we show the separate differential edge-length distributions for the
\ion{H}{ii} regions and diffuse regions. The slopes of the power law
fits are $\alpha=-1.17\pm0.07$ and $\alpha=-1.29\pm0.12$, respectively.
The flattening of the distributions occurs at edge-lengths of about 20 and
32\,pc. These values do not differ substantially, indicating a similarity
in the clustering properties of the \ion{H}{ii} and diffuse regions. 

\subsection{The $\mathrm{H}\alpha$ Flux}
\label{sec4.8}

To make a quantitative estimate of the diffuse emission contribution,
we subtract the contribution of the \ion{H}{ii}
regions from the total $\mathrm{H}\alpha$ flux. 
Since the local background has been removed from the region flux, 
our measurement of the region contribution corresponds to a lower limit.
To obtain an upper limit we subtract the region background fluxes as well.
We find that 35--39\% of the $\mathrm{H}\alpha$ flux in the resolved
frames is attributable to individual \ion{H}{ii} regions. For comparison, 
Ferguson et al.\,(\cite{fer96}) find values of about 50--70\% and
Rozas et al.\,(\cite{roz99}) 40--65\%.
The remaining flux must come from the diffuse regions and background. 
If we determine the same ratio for the smeared frames we obtain a somewhat 
different result. Here 64--74\% of the total emission is found to be
attributable to \ion{H}{ii} regions and only 26--36\% to the diffuse part.

{\it A substantial fraction of the total diffuse emission is included in
the \ion{H}{ii} region luminosities by the effect of smearing.} This
is another indication of the importance of resolution in the
interpretation of \ion{H}{ii} region data.

\section{Discussion}
\label{sec5}

We have shown in the previous sections that the basic properties of 
\ion{H}{ii} regions determined from high resolution HST frames differ 
significantly from even high quality ground based measurements.  
Only a few previous studies have had the necessary linear resolution
to make reliable measurements of their statistical properties, such as 
size and luminosity distributions (Walterbos \& Braun \cite{wal92}; Wyder
et al. \cite{wyd97}).

We simulate the effects of lower resolution on our HST frames by
artificially smearing the data to typical ground based resolutions.
The diameter distribution from high resolution HST observations exhibits a
totally different range of values compared to those determined from lower
resolution data. We find a diameter distribution which is more consistent
with a power law than with the often used exponential function. This puts
into doubt the concept of a characteristic size for \ion{H}{ii} regions.

The luminosity function can be fit by a single slope power law above
a luminosity of $\log\!L=36.8$\,erg/s.  Below this value the distribution
is flat. Unfortunately, the existence of a break at higher luminosities 
can not be reliably determined through direct examination of the
luminosity function derived from the \object{M101} data because of the
small number of regions with luminosities above the expected location of
the break. In order to make such a determination, greater coverage of
\object{M101} will be needed to increase the statistics, particularly at
the high luminosity end. 

The LF of the artificially smeared data (Fig.~\ref{fig06}{\bf a}) is
flatter in the mid-luminosity range. A clear identification of a
high luminosity break can also not be reliably established.
A close comparison of the frames at the two different resolutions
has given us important insight into the influence of resolution on
the measured \ion{H}{ii} region properties.  We find that as the
resolution is reduced regions which are spatially close together are
blended  into a single observed ``region'', along with a significant
amount of diffuse emission.
This process leads to the shifting of regions from low luminosity bins to
higher bins.  A luminosity of $\log\!L\approx38.5$\,erg/s appears to
mark an upper limit for this luminosity shifting effect.  The blending
of regions alone is not sufficient to explain this limiting luminosity,
but requires the inclusion of the diffuse emission.  While still
not fully understood, it appears that the intrinsic properties of
the diffuse emission clouds, such as their size, surface brightness, and
spatial association with the \ion{H}{ii} regions, play an important role.
Modeling of the LF with the  inclusion of  the various
effects of degraded resolution (Sec.~\ref{sec4.1}) would be helpful in
determining the nature of the high luminosity break and its possible
relation to the diffuse emission.

The idea that the high luminosity break is a result of the 
blending of \ion{H}{ii} regions and the masquerading of diffuse regions as
genuine ionization sources differs from the hypothsis that it is due
to a transition in the physical state of the \ion{H}{ii} regions from
ionization bounded to density bounded (Rozas et al. \cite{roz96a};
Beckman et al. \cite{bec98}).
Still a connection might be given by the nature of the diffuse emission. 
If the diffuse emission close to the \ion{H}{ii} regions is only ionized
by the nearest source, then it may be considered as being associated with
the regions. When a region and its associated diffuse emission are thrown
together by the effects of resolution, then the blended region may be seen
as being ionization bounded when sufficient
diffuse gas is included. On the other hand a region may be seen as
density bounded when the amount of diffuse gas included is insufficient
with respect to the luminosity of the internal ionizing source.
In this scenario the amount of diffuse gas included in a blended region is
controlled by the resolution and hence to some extent so would be the
position of the high luminosity break in the \ion{H}{ii} region LF.

With even good ground based data the definition of what is an \ion{H}{ii}
region is heavily influenced by the resolution. At high resolution one can
see that these smeared ``regions'' really consist of a complex of
associated structures, such as diffuse clouds, arclets, shells, and filaments.
Hence the surrounding gas is far from being evenly distributed around the
ionizing source.  With such a non-uniform topology it would be difficult
to achieve a true ionization bounding, because the possibilities for a
Lyman-continuum photon to escape unhindered are numerous. Such a situation
is a poor representation of an idealized Stromgren sphere. It is also not
clear to what degree external sources may contribute to the ionization of
the diffuse component and particularly of the diffuse regions.

A direct examination of a sequence of artificially resolution degraded
frames, and confirmed by the MST analysis, shows that the effects of
smearing become particularly prominent around a critical linear resolution of
about 40\,pc FWHM and thereafter only worsens slowly.
If the high luminosity break in the \ion{H}{ii} region luminosity function
is an artifact of resolution then this characteristic behavior may explain
the stability of the break luminosity over a relatively large range of
galaxy distances. Therefore the stability of the break luminosity
is not unambiguous evidence against blending as a mechanism for 
producing the discontinuity.  Clearly, modeling of this mechanism will be
required to determine its viability.  However, the importance of the
diffuse emission should not be ignored and any model of the blending
will need to take this into account along with the characteristics of
the clustering.

The underlying reason for this behavior seems to be the clustering 
characteristics of the \ion{H}{ii} regions and diffuse emission. The MST
edge-length distribution exhibits a power law form with a flattening at
about 25\,pc. The length at which this transition occurs may be the
primary factor influencing the location of the high luminosity break. The
physical reason for the flattening at this value of edge-length is not clear,
but may be related to a typical maximal fragmentation length within a single
precursor gas-cloud. The power law slope would then represent the distribution
of distances between the precursor gas-clouds. Another possibility could be
related to two different mechanisms of star formation as described by
Scowen et al. (\cite{sco96}). Whereas the power law edge-length distribution
might correspond to the fragmentation of the ISM induced by the passage of a
spiral arm density wave, the flat part may represent typical spatial sizes of
local subsequent star formation. In this case the patches of diffuse emission
might be interpreted as expanding HII region remnants. If the break in the LF
is interpreted as due to these clustering properties, its frequent appearance
derived from ground-based data may suggest a similarity of two different 
regimes of star formation in M101 and other late-type spirals. 
 
There is an indication that the clustering characteristics between
the two observed fields differ.  While the edge-length distributions 
show similar slopes, the value at which  the power law flattens out
differ by about 30\,pc.  It is not clear if this difference
has a physical origin or is merely statistical.
Such a difference is not found when we compare the luminosity
functions or diameter distributions for the different fields. The
flattening of the luminosity function is found for both fields at
$\log\!L=36.8$, which corresponds to a diameter of about 30\,pc.  
If the difference is indeed physical in nature, then one must explain how
the properties of the fragmentation regimes can differ without affecting the
other star formation properties.

Our simple analytic model of the resolution effects on the profiles 
can successfully reproduce the general behavior of the surface
brightness gradients seen at lower resolutions. While the profiles
of the lower luminosity regions ($\log\!L\la38.5$) are dominated by
the point spread function, the profiles of higher luminosity
regions which have spatial sizes greater than the seeing scale are
expected to represent the true properties of the regions. We therefore
interpret the break in the surface brightness gradient distribution as a
resolution effect.  This simple model provides a  prediction on how the
luminosity of the break changes with distance.

The measured central surface brightness of the resolved regions shows
a large scatter and does not exhibit a well defined trend with luminosity.
This is not true for the smeared regions which exhibit a nearly constant
value up to $\log\!L=38.5$\,erg/s. After this the central surface brightness
increases steeply with increasing luminosity.  Our simple model again
reproduces this behavior fairly well and predicts the transition
luminosity will vary with distance.  

In both cases the profile and the central brightness become seeing
dominated at linear resolutions worse than 40\,pc FWHM. Therefore 
this linear resolution marks the minimum resolution required for
reliable determinations of region characteristics.
This kind of resolution is currently only achievable from the ground
for the nearest galaxies. 
Taking this into account the observed breaks at lower resolution do not
seem to be reliable evidence for a transition from ionization bounded to density
bounded \ion{H}{ii} regions.

Concerning the question as to whether the high luminosity break in the
LF can be used as a standard candle as proposed by
Beckman et al. (priv. comm.), more observations will be required 
to validate the stability of the break luminosity with distance
(resolution) and galaxy type.  A clearer understanding of its physical
origin would greatly assist in such an effort. In our interpretation, the
break is a result of the clustering characteristics of the \ion{H}{ii}
regions and diffuse emission. As seen in the MST analysis, the differences
between the two fields we studied in \object{M101} may be an indication
that significant differences in the clustering properties exist even within
a single galaxy.  Depending on the degree to which this were true, the 
utility of a global LF of a galaxy as a measure of its star formation
properties would be compromised.

A further indication of variations in region clustering may be in the
study of the barred galaxy NGC 7479 by Rozas et al. (\cite{roz99}), who
find differences between the bar \ion{H}{ii} region population and that
of the overall disk. Furthermore this galaxy shows a different behavior in
comparison to the non-barred galaxies with respect to the break luminosity
in the central surface brightness distribution. This could be a hint for a
possible variation of \ion{H}{ii} region clustering properties with
galaxy type.

Clearly additional high resolution work is needed to both confirm
and extend these results.  More sophisticated methods should be 
applied to the radial profiles in order to better characterize their
properties. Even a rough determination of cluster age from broad band
imaging would be useful for constraining the physical origin of the
larger scale HI structures and their relationship to the star forming
clouds.

With respect to possible further high resolution observations, 
obtaining better statistics is vital for making further progress on the 
luminosity and diameter distributions. To generalize the results, these
investigations should be carried out for different galaxy types.  
A comparison between a barred and non-barred galaxy would be an
important test of the stability of the high luminosity break as
the dynamics of the two systems are quite different.

\section{Summary}
\label{sec6}

In this paper we present high resolution HST $\mathrm{H}\alpha$ data of
\object{M101}. For a comparison with previous ground based observations
we artificially shift these frames to an effective distance of 20\,Mpc by
rebinning, changing the pixelsize to $0\farcs28$, and convolving with
a seeing gaussian of $0\farcs8$ FWHM.
For both resolutions we derive the integral diameter distribution, 
luminosity function, surface brightness gradients and central surface
brightnesses.

For the integral diameter distribution at high resolution we find a
characteristic diameter $D_0=29.2$\,pc. However, a power law fits the data
better with a slope ${\beta}=-2.84\pm0.16$.
The luminosity function in the resolved case can be fit by a single slope
power law with $\alpha = -1.74$ in the luminosity range between
$10^{36.7}$ and $10^{39.3}$ erg/s. It flattens for lower luminosities.
The surface brightness gradients exhibit a constant limit with a lot of
scatter to steeper gradients. Because of large scatter it is hard to find an
appropriate function of luminosity for the central surface brightness.

Comparing the diameter distribution at the two resolutions, we find a
totally different range of values. Whereas the diameters of the resolved
regions lie between about 10 and 220\,pc, the diameters in the smeared
case show values from about 150\,pc to greater than 400\,pc. We also find a
different functional form at the higher resolution, namely a power law
rather than an exponential function.

The LF in the smeared case exhibits a reduction of regions in lower
luminosity bins, while an excess in the range between $10^{37.8}$
and $10^{38.6}$\,erg/s. We attribute this behavior to the intrinsic
clustering properties of the \ion{H}{ii} regions and diffuse emission
combined with resolution effects. As the resolution is reduced spatially
associated regions are blended together including a significant amount of
diffuse emission. This process leads to a shifting of the regions into
higher luminosity bins and may account for the often observed break in the
LF at $\log\!L=38.6$\,erg/s.

We have further investigated this
effect by examining the clustering quantitatively with a minimal spanning
tree (MST). There we find a power law function for the edge-length
distribution which flattens at about 25\,pc and has an index of about
$\alpha=-1.5$.

The edge-length where the flattening occurs may be the primary influence on
the location of the break. The two different spatial regimes are interpreted
as two different regimes of star formation in M101. If the break in the LF is
interpreted as due to these clustering properties its frequent appearance
derived from ground-based data may suggest a similarity of two different
regimes of star formation in M101 and other late-type spirals.

The surface brightness gradients as a function of luminosity for the
smeared regions may be described by a bilinear fit, but a simple analytic
model shows that this feature results completely from the degraded
resolution. With the same mathematical approach the break of the central
surface brightness of the smeared regions is predicted correctly, indicating
that neither break can be considered as a real physical effect.

The minimum linear resolution needed to reliably measure internal
quantities is found to be about 40\,pc FWHM. Such resolutions are only
currently obtainable from the ground for the nearest galaxies.

In our view the hypothesis that the high luminosity break in the LF is due
to a transition from ionization bounded to density bounded \ion{H}{ii}
regions is still not firmly established and resolution effects need to be
better taken into account when making such studies. The intrinsic clustering
properties of the \ion{H}{ii} regions and the associated diffuse emission
combined with typical resolutions may be able to produce the observed break.  

Further high resolution observations will be needed to confirm these
results and obtain reliable statistics on the characteristics of
\ion{H}{ii} regions in galaxies. Only then will it be possible to
determine whether the clustering characteristics of \ion{H}{ii} regions
differ with galaxy type and environment, and any connection it may have
with the local dynamics.

\begin{acknowledgements}
We wish to thank John Beckman for providing his early results, Rebecca
Koopmann for help with the data reduction, Albert Bosma
for interesting discussions, and Polichronis Papaderos and Uta
Fritze-von Alvensleben for helpful comments. This work was supported in part
by the Deutsche Forschungsgemeinschaft under grant FR 325/39-3.

\end{acknowledgements}

\end{document}